\documentclass[reprint,twocolumn,floats,aip,jcp,english,preprintnumbers,amsmath,amssymb]{revtex4-1}
\usepackage[T1]{fontenc}
\usepackage[latin9]{inputenc}
\usepackage{verbatim}
\usepackage{float}
\usepackage{graphicx}
\usepackage{prettyref}

\makeatletter
\@ifundefined{textcolor}{}
{%
 \definecolor{BLACK}{gray}{0}
 \definecolor{WHITE}{gray}{1}
 \definecolor{RED}{rgb}{1,0,0}
 \definecolor{GREEN}{rgb}{0,1,0}
 \definecolor{BLUE}{rgb}{0,0,1}
 \definecolor{CYAN}{cmyk}{1,0,0,0}
 \definecolor{MAGENTA}{cmyk}{0,1,0,0}
 \definecolor{YELLOW}{cmyk}{0,0,1,0}
 }

\usepackage{dcolumn}
\usepackage{bm}
\usepackage{enumerate}

\makeatother

\usepackage{babel}

\begin{document}

\title{Transferable Pair Potentials for CdS and ZnS Crystals}

\author{Michael Gr{\"u}nwald}
\affiliation{Department of Chemistry, University of California, Berkeley, California 94720}

\author{Phillip L. Geissler}
\affiliation{Department of Chemistry, University of California, and Lawrence Berkeley National Laboratories, Berkeley, California 94720}

\author{Eran Rabani}
\affiliation{School of Chemistry, The Sackler Faculty of Exact Sciences,
  Tel Aviv University, Tel Aviv 69978, Israel}

\date{\today}
\begin{abstract}
A set of interatomic pair potentials is developed for CdS and ZnS
crystals.  We show that a simple energy function, which has been used to describe
the properties of CdSe [{\em J. Chem. Phys.} {\bf 116}, 258 (2002)],
can be parametrized to accurately describe the lattice and elastic constants,
and phonon dispersion relations of bulk CdS and ZnS in the wurtzite and
rocksalt crystal structures. The predicted coexistence pressure of the wurtzite and rocksalt structures, as well as the equation of
state are in good agreement with experimental observations. These new
pair potentials enable the study of a wide range of processes in bulk and nanocrystalline II-VI semiconductor materials.
\end{abstract}
\maketitle

\section{Introduction}
\label{sec:intro}

Many important processes in solid state materials, like the melting transition,\cite{Goldstein92,Dellago2005} structural transformations,\cite{Tolbert95,Grunwald2009}
or diffusion of impurities and defects~\cite{Smit2000} require atomistic resolution in space and time for a comprehensive understanding of the underlying mechanism. Despite major advances in electron microscopy,\cite{Zheng2011} experiments can only provide a coarse-grained view of such processes. Molecular dynamics computer simulations can in principle provide the necessary microscopic perspective, but their predictive power depends on the reliability and feasibility of available models. 

Methods based on first principles that retain a description of the electronic state of the system potentially offer the highest accuracy. Because of their high computational demand, however, they are not currently suited for in-depth studies of systems involving more than a few hundred atoms, or time scales longer than a few tens of picoseconds. Classical interaction potentials, parameterized to reproduce emergent properties of the modeled material, offer a compromise between accuracy and computational speed. Potentials of different functional form and complexity have been developed for materials with widely disparate chemical and physical properties, ranging from water~\cite{WATER} to gold~\cite{Parrinello1988} and biopolymers.\cite{AMBER} Depending on the properties studied, agreement with experiment is usually good, in some cases rivaling or even besting that of affordable \emph{ab initio} methods.

Sparked by a comprehensive experimental study of the structural changes occurring in CdSe nanocrystals under pressure, \cite{Tolbert94,Tolbert95,Alivisatos95,Chen97,Alivisatos00,Jacobs01,Jacobs02} a simple pair potential has been developed~\cite{Rabani02b} and successfully applied to reveal the mechanisms of structural rearrangements in both bulk and nanocrystalline CdSe.\cite{Zahn2005,Grunwald2006,Grunwald2007,Leoni2008,Grunwald2009,Grunwald2009a,Bealing2009,Bealing2010,Martonak2011} However, simulation studies of processes in many other semiconductor materials, or in multi-component systems like core/shell crystals or seeded nanorods, have been precluded by the lack of available models. Here we present a set of model potentials for crystalline CdS and ZnS, two semiconductors with potential use in various light harvesting and
opto-catalytic devices.\cite{Trindade01} The potentials are designed to reproduce the bulk lattice and elastic constants of the relevant crystal structures, as well as phonon dispersion relations. They are specifically constructed to be compatible with each other and with the existing model for CdSe \cite{Rabani02b} and therefore also enable simulations of mixtures of the three compounds.    

The paper is organized as follows: In Section~\ref{sec:pair}, we
discuss the construction of the pair potentials and specify their parameters. 
In Section~\ref{sec:eos}, we apply the models to calculate bulk enthalpies as a function of pressure and the equations of state for
CdSe, CdS and ZnS and compare the predictions to experimental
results. Discussion and conclusions are given in Section~\ref{sec:conclusions}.

\section{The Pair Potential}
\label{sec:pair}
We use the simple model developed for CdSe
\cite{Rabani02b} as a template to also describe CdS and ZnS.
The two-body interatomic potential consists of a long range Coulomb
part and a short range part which is represented by a Lennard-Jones (LJ)
form:
\begin{equation}
V_{ij} = \frac{q_{i} q_{j}}{r_{ij}} + 4 \epsilon_{ij} \left\{
\left(\frac{\sigma_{ij}}{r_{ij}}\right)^{12} -
\left(\frac{\sigma_{ij}}{r_{ij}}\right)^{6} \right\},
\label{eq:Vij}
\end{equation}
where the indexes $i$ and $j$ refer to Cd, Zn, S, and Se atoms. To facilitate transferability and reduce the number of parameters, we
use standard combining rules for interactions of unlike atom types, namely
$\epsilon_{ij}=\sqrt{\epsilon_{i} \epsilon_{j}}$ and $\sigma_{ij} =
\frac{1}{2}(\sigma_{i}+\sigma_{j})$.  

The parameters $q_{i}$, $\epsilon_i$, and $\sigma_i$ were obtained by fitting the lattice and elastic
constants, and phonon dispersion relations of bulk CdS and ZnS in three crystal structures: wurtzite, zinc-blende
and rocksalt. As an additional constraint, the
energy difference between the wurtzite and rocksalt structure at zero pressure
was fitted to {\it ab initio} calculations.\cite{Durandurdu09,Tan11}
This was done to ensure that the wurtzite structure is the
more stable structure at low pressures. The fitting calculations were
performed at $0$~K, although the experimental data were
obtained at finite temperatures.

\begin{table}
\begin{center}
\begin{tabular*}{50mm}{@{\extracolsep{\fill}}r|rrr}
& $q(\mathrm{e})$ & $\sigma \mbox{(\AA)}$ & $\epsilon/k_B \mbox{(K)}$ \\\hline
Zn & 1.18 & 0.02 & 17998.4  \\
S & -1.18 & 4.90 & 16.5  \\
Cd & 1.18 & 1.98 & 16.8  \\
Se & -1.18 & 5.24 & 14.9 \\
\end{tabular*}
\end{center}
\caption[]{Potential parameters defining the
  interatomic interactions in ZnS, CdS and CdSe.}
\label{ta:par}
\end{table}

To obtain transferable potentials and ensure neutrality of the modeled materials, we fixed the magnitude of all ion charges to that of the original CdSe model, \emph{i.~e.}, $|q_i| = 1.18$~e. We then proceeded in the following way. To obtain a model for CdS compatible with that of CdSe, we relaxed the LJ parameters for sulfur, keeping parameters for Cd fixed. In the second step, to arrive at a model for ZnS, we likewise relaxed the LJ parameters for Zn, keeping S fixed. Thus, a total of four parameters ($\sigma_{Zn}$, $\sigma_S$, $\epsilon_{Zn}$, and $\epsilon_S$) was determined with the fitting.

We used a relaxed fitting procedure similar to the one discussed
in Refs. \onlinecite{Gale96} and \onlinecite{Gale97}.  This procedure
is substantially more expensive computationally than the conventional
fitting.  However, it allows a much higher quality of fitting which is
required to properly reproduce the structural properties of CdSe.  In
the relaxed fitting procedure, the error was defined on the residual
of the structural and dynamical properties of the optimized
configurations of the different crystal phases rather than on the
experimental observed structures.  Namely, the configurations and the
lattice constants of each crystal phase were quenched using the
conjugate gradient algorithm, and the aforementioned properties were
calculated and compared to the experimental values for the quenched
structures. In all the results reported here we have used Ewald sums to evaluate the electrostatic interactions, \cite{Frenkel2002}
with a partitioning parameter between the two spaces chosen to
minimize the computational effort.\cite{Allen87} The Lennard-Jones part of the potential was cut at half the box length ($\approx 10$~\AA{}).

\begin{figure}
\includegraphics[width=1\columnwidth]{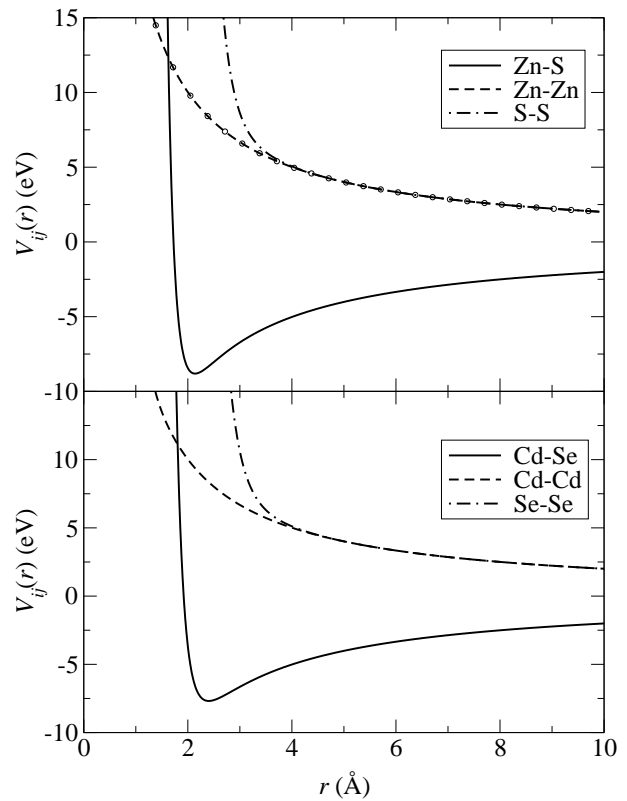}
\caption{Interatomic pair potentials for Zn-Zn, S-S, and
  Zn-S (top panel) and Cd-Cd, Se-Se and Cd-Se (bottom panel). Open
  circles in the upper panel show the pure Coulomb repulsion term.}
\label{fig:pot}
\end{figure}

The final parameter values resulting from the fit (including those for CdSe from
Ref.~\onlinecite{Rabani02b}) are summarized in Table~\ref{ta:par}.
Consistent with their negative charge, the van der Waals (vdW) radii of the anions S and Se are significantly larger than those of the cations. In agreement with the corresponding ionic radii, the vdW radius of S is smaller than that of Se and the radius of Zn is smaller than that of Cd. For Zn we find that the best fit is obtained for a nearly vanishing vdW radius and a large value of $\epsilon$. These particular values are a result of the constraints imposed on the charge of the ions and the sequence of fitting steps. Because of the small value of $\sigma$, the forces between Zn atoms at relevant distances are governed by the Coulomb term only; the large value of $\epsilon$ sets the Zn-S bond length. A plot of the interatomic pair potentials is shown in Fig.~\ref{fig:pot}.

\begin{figure}
\includegraphics[width=1\columnwidth]{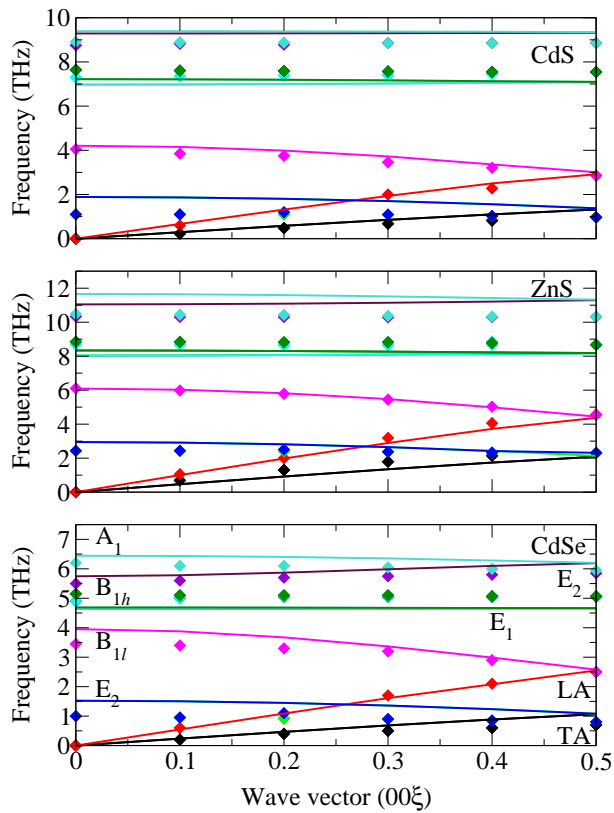}
\caption{Phonon dispersion relations of wurtzite CdS (upper panel),
  ZnS (middle panel), and CdSe (lower panel) along the $\Gamma A$
  direction. The filled diamonds represent literature results.\cite{Cheng09,Cardona99,Birman70}}
\label{fig:phonon}
\end{figure}

The accuracy of the model in reproducing the phonon dispersion
relations of wurtzite ZnS, CdS and CdSe along the $\Gamma A$ direction
is shown in Fig.~\ref{fig:phonon}. The experimental results were
obtained from inelastic neutron scattering~\cite{Cardona99} on
$^{116}\mbox{CdSe}$ and Raman scattering for ZnS.\cite{Cheng09} For
CdS, we have compared our results with calculations providing good
agreement with infrared absorption.\cite{Birman70} The split between
the $\mbox{E}_{2}$ and $\mbox{B}_{1l}$ at the $\Gamma$-point indicates
the ionic nature of the material.\cite{Dixon76} The overall
frequencies in ZnS are higher due to the lighter masses of the atoms
compared to CdS and CdSe. The TA and LA branches are back-folded into
the lower $\mbox{E}_{2}$ and $\mbox{B}_{1l}$ branches, respectively.
Similar back-folding occurs for the $\mbox{A}_{1}$, $\mbox{B}_{1h}$ and
the upper $\mbox{E}_{2}$ and $\mbox{E}_{1}$ branches.  As can
be clearly seen in the figure, our simplified model captures the back-folding
in all branches. The overall agreement between the calculations and
the experimental results is reasonable given the simple
form of the potential (cf. Eq.~(\ref{eq:Vij})).  The
model performs slightly better for ZnS and is more accurate for the
lower frequency branches.  The agreement can likely be improved using a
polarizable model, as is well known for alkali
halides.\cite{Birman70,Cochran71} We note that even better agreement with experiment
has been achieved with {\it ab initio} methods~\cite{Cardona99,Kushwaha10}.

In Table~\ref{ta:constants} we compare the lattice
and elastic constants calculated using our model with the
corresponding experimental values.  Note that the results for CdSe
differ slightly from our original report.\cite{Rabani02b} This deviation is a
consequence of using a smaller tolerance for the minimization, which was considered excessively
cumbersome at the time the CdSe parameters were generated. The calculated
lattice constants are within $1\%$ of the experimental values, except
for the case of the rocksalt structure in ZnS, where the error is
$3\%$.  The agreement between the calculated elastic constants and the
experimental values is qualitative, with small errors in $C_{11}$ and
$C_{44}$.

To compare the bulk modulus obtained from our model with
experimental results we have used the well known relation between the
bulk modulus and the elastic constants:
\begin{equation}
B = \frac{C_{33} (C_{11} + C_{12}) - 2C_{13}^{2}}{C_{11} + C_{12} + 2C_{33}
- 4C_{13}},
\label{eq:Bhex}
\end{equation}
for hexagonal symmetry and
\begin{equation}
B = \left[(C_{11} + C_{22} + C_{33})/3 + 2C_{12} \right]/3,
\label{eq:Bcub}
\end{equation}
for cubic symmetry.  The calculated bulk moduli for wurtzite ZnS, CdS
and CdSe are in reasonable agreement with the experimental values
(which show some spread from one report to the other). Moreover,
the values of the bulk modulus are comparable in accuracy to those
obtained by {\em ab-initio} methods.\cite{Durandurdu09,Tan11}

\begin{table}
\begin{center}
\begin{tabular*}{80mm}{@{\extracolsep{\fill}}c|ccc}
CdS & Wurtzite & Zinc-blende & Rocksalt\\\hline
$a$ & 4.16 (4.13)~\cite{Hellwege82a,Hellwege82b} & 5.83 (5.82)~\cite{Wyckoff63} & 5.43 (5.44)~\cite{Owen63}\\
$c$ & 6.61 (6.70)~\cite{Hellwege82a,Hellwege82b} &              & \\
$C_{11}$ & 107.3 (90.7) & 99.6  & 96.9 \\
$C_{12}$ & 35.8  (52.1) & 32.9  & 53.7 \\
$C_{13}$ & 15.9  (51.0) &       &      \\
$C_{33}$ & 144.3 (93.8) &       &      \\
$C_{44}$ & 19.1  (15.0) & 41.3  & 53.8 \\
$C_{66}$ & 20.1  (16.3) &       &      \\
$B$     & 54.0  (65.0) & 55.1 (55.0) & 68.1\\ \\

ZnS & Wurtzite & Zinc-blende & Rocksalt\\\hline
$a$ & 3.89 (3.85)~\cite{Durandurdu09,Wright95} & 5.48 (5.41)~\cite{Durandurdu09,Wright95} & 5.20 (5.06)~\cite{Durandurdu09,Wright95}\\
$c$ & 6.26 (6.29)~\cite{Durandurdu09,Wright95} &                     & \\
$C_{11}$ & 161.4 (131.2) & 150.1 (94.2) & 109.7\\
$C_{12}$ & 53.8  (66.3) & 51.4  (56.8) & 88.0\\
$C_{13}$ & 28.2  (50.9) &      &     \\
$C_{33}$ & 213.1 (140.8) &      &     \\
$C_{44}$ & 32.4  (28.6) & 62.2 (43.6) & 88.1\\
$C_{66}$ & 53.8  (32.4) &      &     \\
$B$     & 82.3  (82.1) & 84.3 (69.3) & 94.6\\ \\

CdSe & Wurtzite & Zinc-blende & Rocksalt\\\hline
$a$ & 4.37 (4.30)~\cite{Cohen88} & 6.13 (6.08)~\cite{Cohen88} & 5.74 (5.71)~\cite{Alivisatos00}\\
$c$ & 6.97 (7.01)~\cite{Cohen88} &                     & \\
$C_{11}$ & 87.2  (74.6) & 82.4 & 73.2\\
$C_{12}$ & 29.1  (46.1) & 26.6 & 45.3\\
$C_{13}$ & 13.3  (39.4) &      &     \\
$C_{33}$ & 118.0 (81.7) &      &     \\
$C_{44}$ & 15.9  (13.0) & 33.4 & 45.3\\
$C_{66}$ & 29.1  (14.3) &      &     \\
$B$     & 44.1  (53.4) & 45.1 & 54.6\\
\end{tabular*}
\end{center}
\caption{Calculated lattice constants (in \AA{}), elastic constants (in GPa), and
  bulk modulus (in GPa) of CdS, ZnS and CdSe for three crystal
  structures. Experimental results are given in parenthesis when
  available. (Elastic constants and bulk moduli were taken from Refs.~\onlinecite{Wright95,Bonello93,Jaffe63,Corll67}.)}
\label{ta:constants}
\end{table}

\section{Phase Diagram and Equation of State}
\label{sec:eos}
\begin{figure}
\includegraphics[width=0.9\columnwidth]{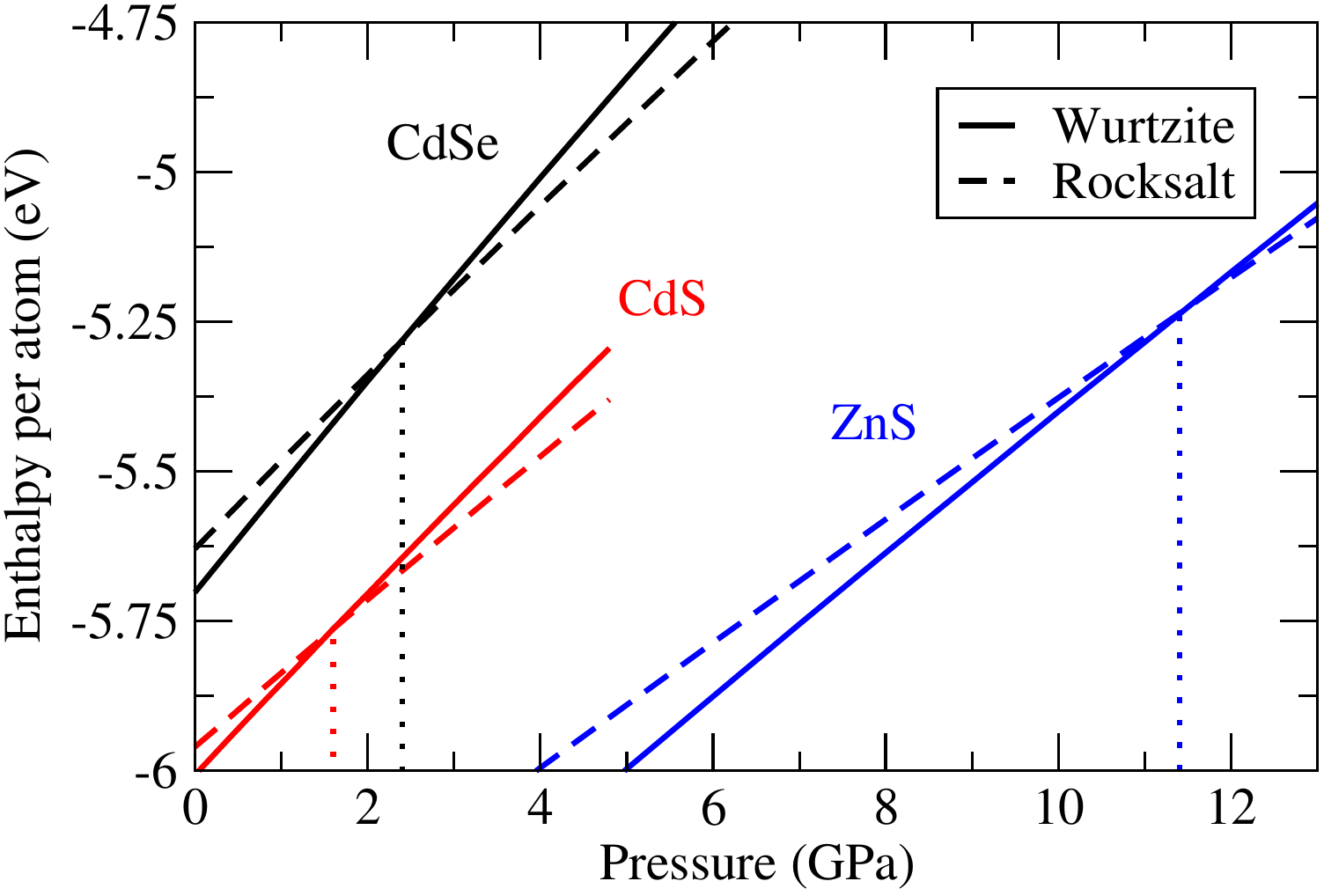}
\caption{Enthalpies as a function of pressure of bulk ZnS (upper
  panel), CdS (middle panel), and CdSe (lower panel), at a temperature of 300~K. The solid and
  dashed lines show results obtained for wurtzite and
  rocksalt crystal structures, respectively.  The vertical dotted lines mark points of equal enthalpies and approximately correspond to thermodynamic
wurtzite to rocksalt transition pressures in the respective materials.}
\label{fig:phase}
\end{figure}
To test the accuracy of our models on quantities not directly used in the fitting procedure, we calculated coexistence pressures for the wurtzite and rocksalt structures at $T=300$~K, as well as equations of state for all three crystal structures.

In Fig.~\ref{fig:phase} we plot the enthalpy as a function of pressure for bulk
ZnS, CdS, and CdSe.  The pressure was varied between $0$ and $15$ GPa
using the constant pressure Monte Carlo simulation
technique.\cite{Frenkel2002} For each crystal structure, we have used a periodically replicated simulation box of more
than $400$ atoms, and averaged the results over $50,000$ Monte Carlo cycles. Each cycle on average consisted
of one attempted displacement move for all atoms, and one attempted change of the simulation box volume.

Approximating the true coexistence pressure by points of equal enthalpy, we find that the phase transformation from 
wurtzite to the rocksalt structure occurs at $\approx 2.4~\mbox{GPa}$, $\approx 1.6~\mbox{GPa}$, and $\approx
11.4~\mbox{GPa}$ for CdSe, CdS and ZnS, respectively.  These values agree well with experimentally observed transition pressures of $\approx
2.5~\mbox{GPa}$,\cite{Mariano63,Alivisatos00} $\approx
2.5-3.2~\mbox{GPa}$,\cite{Edwards61,Samara62,Venkateswaran85,Zhao89} and $\approx
12~\mbox{GPa}$,\cite{Desgreniers00} for CdSe, CdS, and ZnS, respectively.
In all three materials, the zinc-blende crystal structure is not stable in the pressure range studied here; its enthalpy (not shown) is slightly higher than the corresponding wurtzite enthalpy.

\begin{figure}
\includegraphics[width=1\columnwidth]{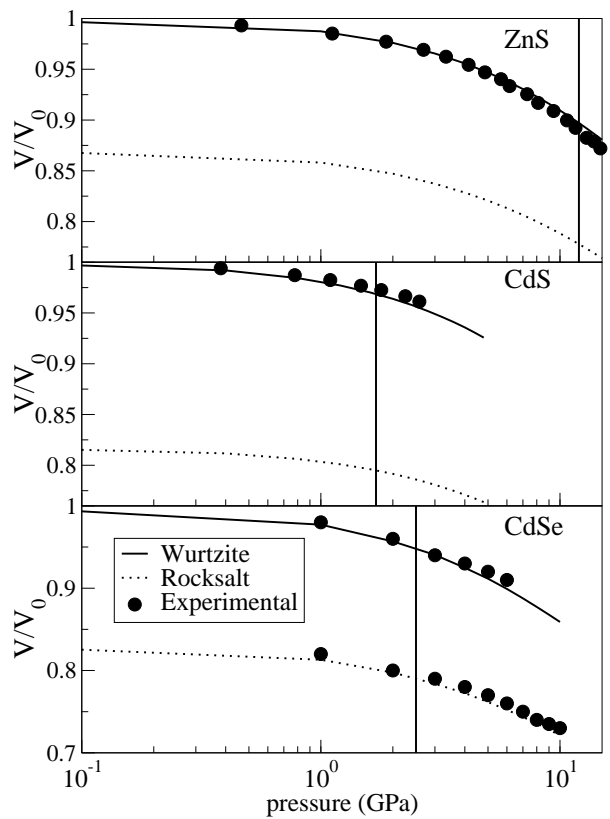}
\caption{Volume as a function of pressure for bulk ZnS
  (upper panel), CdS (middle panel), and CdSe (lower panel).  The
  solid and dotted lines are the calculated results for wurtzite and
  rocksalt crystal structures, respectively. $V_{0}$ is the unit cell volume of the wurtzite structure at zero
  pressure. Filled circles show experimental results.\cite{Alivisatos00,Sowa05,Desgreniers00}
 (Note that the CdSe data were obtained in experiments on $45$~{\AA} diameter CdSe nanocrystals.\cite{Alivisatos00})}
\label{fig:eos}
\end{figure}

In Fig.~\ref{fig:eos} we plot the equation of state (volume as a function of pressure) for all three
materials. We find excellent agreement with experiments on CdSe ,\cite{Alivisatos00} CdS,\cite{Sowa05} and
ZnS.\cite{Desgreniers00} 

\section{Conclusions}
\label{sec:conclusions}
We have developed a set of transferable pair potentials for CdS and
ZnS whose form is similar to that used for CdSe.\cite{Rabani02b} The
model consists of positively and negatively charged ions (Cd/Zn and
S/Se, respectively) which interact via a Coulomb potential,
supplemented by short-range repulsion terms and van der Waals
attractive terms. In order to be able to model alloys and hetero-structures of
CdSe/CdS/ZnS, we used standard combining rules for the cross
terms and fixed the magnitude of the charges to the value obtained for
CdSe,\cite{Rabani02b} thereby reducing
the total number of fitting parameters to $4$.  The
parameters were fitted to reproduce lattice and elastic constants, and
phonon dispersion relations of wurtzite, zinc-blende and rocksalt
crystals structures.

We have calculated the transition pressure of the wurtzite to rocksalt transformation, as well as equations of state at room temperature for all three materials.  Our results are in good
agreement with experiments, thus verifying the accuracy and
practicability of the pair potential on quantities not used
in the fitting procedure.

As a final note, we point out that the simple functional form of the potential naturally limits the portability of our model. In particular, we have tested its accuracy in reproducing the lattice constants of ZnSe, a material whose properties were not included in the fit but which can be easily modeled using the parameters for Zn and Se. We find that deviations from experimental values are on the order of $5$\%, considerably larger than for the other materials.

The current work extends our previous work on CdSe and provides a
basis for the study of structural properties and dynamical processes of a larger variety of
physically interesting materials. These include phase transformation
in core-shell structures, alloys which are important for suppression
of the Auger process,\cite{Efros10,Lifshitz11} seeded rods, and
more.

\section{Acknowledgments}
MG was supported by the Austrian Science Fund (FWF) under Grant number
J3106-N16.  ER thanks the FP7 Marie Curie IOF project HJSC and the
Miller Institute for Basic Research in Science at UC Berkeley for
financial support via a Visiting Miller Professorship.


\end{document}